\documentstyle[12pt]{article}
\font\titlefont=cmbx10 scaled \magstep3

\def\lprox{\mathrel{\raise .3ex\hbox{$<$\kern-
.75em\lower1ex\hbox{$\sim$}}}}

\begin{document}

\begin{flushright}
\vspace*{-2cm}
TUTP-93-2 \\ March 1993
\vspace*{2cm}
\end{flushright}

\begin{center}
{\titlefont MOTION OF INERTIAL OBSERVERS THROUGH NEGATIVE ENERGY}\\
\vskip .7in
L.H. Ford \\
\vskip .2in
Institute of Cosmology\\
Department of Physics and Astronomy\\
Tufts University\\
Medford, Massachusetts 02155\\
\vskip .5in
Thomas A. Roman
\vskip .2in
Department of Physics and Earth Sciences\\
Central Connecticut State University\\
New Britain, Connecticut 06050\\
\end{center}

\newpage
\begin{abstract}
      Recent research has indicated that negative energy fluxes due to quantum
coherence effects obey uncertainty principle-type inequalities of the form
$|\Delta E|\,{\Delta \tau} \lprox 1\,$. Here $|\Delta E|$ is the magnitude
of the negative energy which is transmitted on a timescale $\Delta \tau$. Our
main focus in this paper is on negative energy fluxes which are produced by the
motion of observers through static negative energy regions. We find that
although a quantum inequality appears to be satisfied for radially moving
geodesic observers in two and four-dimensional black hole spacetimes, an
observer orbiting close to a black hole will see a constant negative energy
flux. In addition, we show that inertial observers moving slowly through the
Casimir vacuum can achieve arbitrarily large violations of the inequality. It
seems likely that, in general, these types of negative energy fluxes are not
constrained by inequalities on the magnitude and duration of the flux. We
construct a model of a non-gravitational stress-energy detector, which is
rapidly switched on and off, and discuss the strengths and weaknesses of
such a detector.

\end{abstract}
\newpage

\baselineskip=24pt
\section{Introduction}
\label{sec:intro}
       It has been known for some time that quantum field theory allows
violations of the weak energy condition (WEC) \cite{HE} in the form of locally
negative energy densities and fluxes \cite{EGJ}. A classic example is the
Casimir effect \cite{C,MB}. However, if the laws of physics place no
restrictions on such violations, then it would be possible to use
negative energy to produce gross macroscopic effects. These effects
might include violations of the second law of thermodynamics \cite{F78},
causality \cite{MTU}, and cosmic censorship \cite{FR1,FR2}.

  Fortunately, quantum field theory does impose some restrictions on the
extent of WEC breakdown. One such restriction takes the form of an average
of the WEC taken over a null geodesic \cite{T,B,TR,K,WU}. A version
of this ``averaged weak energy condition'' (AWEC) holds for a quantized scalar
field for physically reasonable quantum states in a wide variety of
two and four-dimensional spacetimes \cite{K,WU}. However, in general it
does not appear to hold in an arbitrary curved four-dimensional spacetime
\cite{WU}. Other constraints on WEC violation are uncertainty principle-type
inequalities on the magnitude and duration of negative energy fluxes due to
quantum coherence effects. Such ``quantum inequalities'' are satisfied by
this type of negative energy flux, as seen by inertial observers in two and
four-dimensional flat spacetime \cite{F78,F91}. A typical form of this
kind of inequality (in two spacetime dimensions) is
\begin{equation}
|F|\,(\Delta\tau)^2\lprox 1\,,               \label{eq:QI}
\end{equation}
where $|F|$ is the magnitude of the negative energy flux and $\Delta\tau$
is its duration.  Although
they are not covariantly formulated, in at
least some cases, constraints of the form Eq. (~\ref{eq:QI}) are
stronger than those provided by AWEC \cite{FR2}. A similar inequality was
found to hold for a quantized massless,
minimally-coupled scalar field propagating on two and four-dimensional
extreme Reissner-Nordstr{\o}m black hole backgrounds. For these cases,
it was shown that the magnitude of the change in the mass of the black hole
$|\Delta M|$, due to the absorption of an injected negative energy flux,
and the effective lifetime, $\Delta T$, of the naked singularity thus
produced were limited by a quantum inequality of the form:
\begin{equation}
|\Delta M|\,{\Delta T} \lprox 1\,.     \label{eq:DMDT}
\end{equation}
The constraint given by Eq. (~\ref{eq:DMDT}) prevents an unambiguous
observation of a violation of cosmic censorship \cite{FR1,FR2}.

Although quantum inequalities prevent the manipulation of negative
energy to produce gross macroscopic effects, such as violations of
the second law, they do not prevent the detection of negative
energy \cite{FGO}. However, recent calculations \cite{KF} indicate
that negative energy densities in flat spacetime are subject to large
fluctuations. This suggests that the semiclassical theory of gravity
may not always yield an accurate description of the gravitational
effects of negative energy.

        The purpose of the present paper is to further explore the generality
of these quantum inequality-type constraints on negative energy fluxes.
An apparent counterexample
is the case of a static observer near the horizon of an evaporating
black hole. The stress tensor of a quantized field in the frame of reference
of such an observer corresponds to a constant negative energy flux directed
into the black hole. (Here we are neglecting the backreaction of the Hawking
radiation on the black hole.) One might attempt to interpret this ingoing
negative flux as really being an outgoing positive flux. However, the energy
density in this frame is negative, so it seems more natural to regard it
as ingoing negative energy. A similar apparent counterexample was suggested
by Davies \cite{Davies}. A mirror in a two-dimensional spacetime which
moves to the right
with increasing acceleration will radiate a negative energy flux to the
right \cite{FD}. The energy flux in the frame of reference of a
(noninertial) observer who runs ahead of this mirror can be constant
and negative. An inertial observer collides with the mirror in a finite time.

Ottewill and Takagi \cite{OT} have also
shown that an observer who co-moves with a mirror which is slowly lowered
in the vicinity of a black hole sees a net negative energy flux. They showed
that the sign of this flux is directly due to the negativity of the static
vacuum energy density swept through by the comoving observer. By lowering the
mirror arbitrarily slowly, they found that one could obtain an arbitrarily
large
violation of the flux inequality.

      However, these are all noninertial observers.  It is well
known that noninertial observers see Unruh radiation effects \cite{Unruh}.
An accelerated detector in flat spacetime responds as if it were immersed
in a thermal bath of particles, even though the stress tensor is zero.
Similarly, a detector suspended near the horizon of an evaporating black
hole behaves as if it were surrounded by a positive energy thermal bath
\cite{Unruh,UW}. Thus the Unruh effect dominates the effects of the negative
energy density and flux. Therefore it appears unlikely that such observers
would be able to use these negative energy fluxes to produce
violations of the second law of thermodynamics. It would be quite surprising
if it were otherwise,
since the ingoing negative energy flux is required to {\it maintain}
consistency with the generalized second law. Inertial observers do not
encounter any such acceleration radiation
which could otherwise mask the effects of the negative energy. Hence in
this paper, we restrict our attention to geodesic observers.

In Sec. ~\ref{sec:CE} we consider
the effective negative energy flux seen by an inertial observer moving
through the  Casimir vacuum. We show that arbitrarily large violations of
the flux inequality, Eq. (~\ref{eq:QI}), can be obtained for slowly moving
observers and argue that such violations are likely to be generic for
observers who move slowly through more general static negative energy
regions. In Sec. ~\ref{sec:obh}, we discuss geodesic observers in black hole
spacetimes.  In a numerical analysis,
we show that for observers moving along radial geodesics in two and
four-dimensional black hole spacetimes, the quantum inequalities
appear to be obeyed. However, geodesic observers in circular orbits
near a black hole may experience a constant negative energy flux,
and thus an arbitrarily large violation of the quantum inequalities.

  In an attempt to better understand the physical implications of these
violations, we present a model of a non-gravitational
stress-energy detector in Sec. ~\ref{sec:detect}. We find that one can
construct a local energy density detector, but that this detector appears
to require rapid switching. Consequently, such detectors do not respond
to the cumulative effects of a flux of negative energy. We discuss the
difficulties inherent in the construction of a non-gravitational energy
flux detector which would register such cumulative effects.
A summary of our results is contained in Sec.~\ref{sec:summary}.
Unless otherwise noted, units with $G=\hbar=c=1$ will be used.

\section{Motion Through Casimir Energy}
\label{sec:CE}

     We wish to consider ``effective'' negative energy fluxes produced by
the motion of observers through  static negative energy backgrounds.
    Let $u^\mu$ be the four-velocity (or two-velocity in the case of
two spacetime dimensions) of an inertial observer. The energy density
and flux in this observer's frame are given by
\begin{equation}
U = T_{\mu\nu} {u^\mu} {u^\nu},    \label{eq:enden}
\end{equation}
and
\begin{equation}
F = - T_{\mu\nu} {u^\mu} {n^\nu},    \label{eq:flux}
\end{equation}
respectively. Here $n^\mu$ is a spacelike unit normal vector, for which
\begin{equation}
{n_\mu}{u^\mu} = 0.  \label{eq:orthog}
\end{equation}

     Consider the following simple scenario for the generation of an
effective negative energy flux. If we take a quantized scalar field in a
two-dimensional Minkowski spacetime and topologically identify the spatial
dimension, then the ground state of the field will be the Casimir vacuum state.
The vacuum expectation values of the stress-energy tensor in this state are
given by \cite{BD}:
\begin{equation}
\rho=\langle T_{tt}\rangle =-{\pi\over{6L^2}}     \label{eq:CTtt}
\end{equation}
\begin{equation}
p= \langle T_{xx}\rangle =-{\pi\over{6L^2}}\,,     \label{eq:CTxx}
\end{equation}
where $L$ is the periodicity length. The closure of the spatial dimension
introduces a preferred reference frame.  Let us now consider an inertial
observer who moves with velocity $v$ relative to this frame so that
$u^\mu = \gamma(1,v)$ and $n^\mu = -\gamma(v,1)$, where
$\gamma=(1-v^2)^{-1/2}$.
This observer will see an effective negative energy flux given by
\begin{equation}
      F  = v{\gamma}^2\,(\rho+p)=-v{\gamma}^2\biggl({\pi\over{3L^2}}
\biggr)\,.                              \label{eq:Cflux}
\end{equation}
Here we have chosen the sign of $n^\mu$ so that $F>0$ when the energy density
and pressure are both positive. Let
\begin{equation}
\tau=\eta \,\biggl({L'\over v}\biggr)=\eta \,\biggl({L\over {v\gamma}}\biggr)
             \label{tau}
\end{equation}
be the proper time for the observer to traverse some fraction $\eta$
of the closed space, where $L'=L/\gamma$ is the periodicity length in the
moving observer's frame. Then from the above
expressions it is easily shown that
\begin{equation}
|F|\,{\tau}^2 = \eta^2\biggl({\pi\over 3v}\biggr)\,,     \label{eq:CQI}
\end{equation}
where $|F|$ is the magnitude of the flux.
Note that by making $v$ arbitrarily {\it small} we can make the right-hand
side of Eq. (~\ref{eq:CQI}) arbitrarily large compared to one, and hence
produce an arbitrarily large violation of the flux inequality. This
violation at small velocities was first noted by Ottewill and Takagi
\cite{OT}.

   Klinkhammer
has used this same spacetime and quantum state to construct a violation
of the averaged weak energy condition \cite{K}. An important feature of
his example is the periodicity of the space, because causal curves can
repeatedly traverse the same negative energy region.
By contrast, our counterexample does not depend on this feature.
Note that in our example, to achieve a violation of the flux inequality
the observer is not required to make even one complete traversal of the space.
A similar result to Eq. (~\ref{eq:CQI}) is also true for plate-type boundary
conditions in both two and four dimensions. By using the methods of
Sec. ~\ref{sec:obh},
one can show that an analogous violation of the flux inequality occurs for an
observer falling into a static ``zero-tidal-force'' Morris-Thorne
\cite{MT} traversable wormhole \cite{TRU}.

    Such violations should
occur quite generally whenever an observer moves slowly through a region of
static negative energy. For example, in a general two-dimensional case let
the characteristic proper size of the region be $L$ and the magnitude of the
negative energy be of order $L^{-1}$, as is the case in the Casimir effect.
Then on dimensional grounds,
the flux should $\sim v/{L^2}$, while the time taken to traverse the
region is of order $L/v$. Therefore, $|F|{\tau}^2\sim 1/v$.

      Although the flux inequality may be violated, this does not
necessarily mean that there are no restrictions on the negative energy.
Returning to our two-dimensional Casimir example, let the magnitude of
the total vacuum energy over the region $L$ be $|E|$. Then from
Eq. (~\ref{eq:CTtt}), we see that
\begin{equation}
|E|\,L= |\rho| \,L^2=\bigl({\pi\over 6}\bigr)\,.
\end{equation}
In the moving observer's frame:
\begin{equation}
|E'| \,L'= |\rho'| \,{L'}^2={\pi\over 6}\bigl(1+v^2 \bigr)\,.  \label{CE'L'}
\end{equation}
Since $v<1$, it follows that $|E'|\,L'<\pi/3$.
Thus, at least for the case of the
two-dimensional Casimir effect, there seems to be a restriction on the
total negative energy and the size of the region occupied by it.
This constraint is of the form originally suggested in Ref. \cite{F91}.
However, in four-dimensions the issue is less clear due to the presence
of additional length scales, e.g. the transverse dimensions of a pair of
Casimir plates. It is also possible that even in two-dimensions the
constraint might be circumvented due to the contributions by other additional
fields to the vacuum energy.

\section{Observers near Black Holes}
\label{sec:obh}

   In this subsection we consider inertial observers moving
in two and four-dimensional
black hole backgrounds. We will calculate the energy density and flux
in the frame of such observers in order to check whether quantum inequalities
of the form of Eq. (~\ref{eq:QI}) are satisfied.

\subsection{Radially Moving Observers}

In this section, we treat observers which are moving radially toward or
away from a Schwarzschild black hole.
First consider a two-dimensional Schwarzschild black hole, where
closed form expressions for $T_{\mu\nu}$ are available. We understand
$T_{\mu\nu}$ to denote a quantum expectation
value in a specified vacuum state. The metric is
\begin{equation}
ds^2 = -C dt^2 + C^{-1} dr^2,
\end{equation}
where $C = 1 - 2M/r$. The geodesic equations imply that
\begin{equation}
{{dr} \over {d\tau}} = \pm \sqrt{k^2 -C}. \label{eq:geodeq}
\end{equation}
A geodesic observer's two-velocity is
\begin{equation}
u^\mu = (u^t, u^r) = \Bigl({{dt} \over {d\tau}}, {dr\over {d\tau}} \Bigr) =
        \Bigl( {k \over C}, \pm \sqrt{k^2 -C} \Bigr). \label{eq:umu}
\end{equation}
Here the $+$ and $-$ refer to outgoing and ingoing observers, respectively.
The constant $k$ is the energy per unit rest mass; $k \geq 1$ corresponds to
unbound trajectories whereas $k<1$ corresponds to bound trajectories. In
the latter case, a particle is constrained to move in the region where
$k^2 \geq C.$ From Eqs. (~\ref{eq:orthog}), (~\ref{eq:umu}) and the fact that
$n^\mu$ is a unit vector, we have that
\begin{equation}
n^\mu = (n^t, n^r) =  \Bigl( C^{-1} \sqrt{k^2 -C}, k \Bigr), \label{eq:no}
\end{equation}
for an outgoing observer, and
\begin{equation}
n^\mu = (n^t, n^r) =  \Bigl( -C^{-1} \sqrt{k^2 -C}, k \Bigr)   \label{eq:ni}
\end{equation}
for an ingoing observer. In both cases, the sign of $n^r$ is chosen so that
the Hawking flux at infinity is positive.

     In our two-dimensional discussion, we will consider observers moving
in both the Unruh and Boulware vacua. The stress tensor components in the
Unruh vacuum are \cite{Unruh77}:
\begin{equation}
T_{tt}={1\over{24\pi}} \biggl({{7M^2}\over{r^4}}-{{4M}\over{r^3}}
       +{1\over{32M^2}}\biggr)\,,    \label{eq:UTtt}
\end{equation}
\begin{equation}
T_{tr}= -{1\over{24\pi}} \biggl(1-{{2M}\over r}\biggr)^{-1}{1\over{32M^2}}\,,
\label{eq:UTtr}
\end{equation}
and
\begin{equation}
T_{rr}=-{1\over{24\pi}} \biggl(1-{{2M}\over r}\biggr)^{-2}
        \biggl({{M^2}\over{r^4}}-{1\over{32M^2}}\biggr)\,. \label{eq:UTrr}
\end{equation}
The corresponding components in the Boulware vacuum are given by:
\begin{equation}
T_{tt}={1\over{24\pi}} \biggl({{7M^2}\over{r^4}}-{{4M}\over{r^3}}
       \biggr)\,, \label{eq:BTtt}
\end{equation}
\begin{equation}
T_{tr}=0\,,  \label{eq:BTtr}
\end{equation}
and
\begin{equation}
T_{rr}=-{1\over{24\pi}}{{M^2}\over{r^4}} \biggl(1-{{2M}\over r}\biggr)^{-2}\,.
\label{eq:BTrr}
\end{equation}
The energy density and flux can be obtained by substituting the above
expressions into Eqs. (~\ref{eq:enden}) and (~\ref{eq:flux}).

     We want to be able to distinguish a ``genuine'' negative energy flux
moving to the right from a positive flux moving to the left. Hence we will
consider the flux to be negative when the energy density, $U$, is
simultaneously
negative. Our goal is to determine whether quantum inequalities of the form
Eq. (~\ref{eq:QI}) hold for geodesic observers in these quantum states. The
strategy is the following. First specify the geodesic as ingoing or outgoing
and choose a value of $k$. Next, numerically determine the range in $r$ over
which $U<0$ for that observer by using Eq. (~\ref{eq:enden}) and the
appropriate
components of the stress tensor. We then numerically integrate the geodesic
equation, Eq. (~\ref{eq:geodeq}), to find $\tau(r)$. The flux as a function
of $r$ is constructed using Eq. (~\ref{eq:flux}) and the appropriate
stress tensor components. Finally, $|F(r)|$ is graphed as a function of
$\tau(r)$ to yield $|F(\tau)|$.

    In two dimensions, we can evaluate the relevant expressions for
all values of $r$ up to $r=0$. A representative example of a graph of $|F|$
as a function of $\tau$ is given in Fig. (1), for $k=1$ in the Unruh vacuum.
This value of $k$
corresponds to free fall from rest at infinity. Using Eq. (~\ref{eq:enden}),
a numerical computation shows that $U<0$ and $F<0$ for this observer in the
range $0 \leq r \lprox 0.36M$. Note that although the flux diverges as the
observer approaches the singularity at $r=0$, most of the negative energy
flux is emitted in a very small fraction of the total time interval shown in
the graph. It therefore seems reasonable to consider the
time interval $\Delta\tau$ of Eq. (~\ref{eq:QI}) not to be the entire
time over which the flux is negative, but rather a timescale which
characterizes the change in the flux. For example, in Fig. (1) we see that
$|F|$ increases more rapidly as $\tau \approx 0.1$ is approached.
We could take the characteristic time to be that over which the flux changes by
a factor of $2$ or $3$. This time will decrease as $|F|$ increases. However,
near the maximum of $|F|$ in Fig. (1), we have $\Delta\tau \lprox 0.01\,M$.
The maximum value
of $|F|$ shown in the graph is less than about $150\,M^{-2}$ and hence
$|F|(\Delta\tau)^2 \lprox 0.01$. From the graph it can be seen that
as the singularity is approached, this timescale rapidly decreases as
the flux increases, such that $|F|(\Delta\tau)^2 \lprox 1$. We have performed
the same analysis for the following values of $k$ for infalling observers in
the Unruh vacuum: $0.001,\,0.1,\,0.5,\,2,\,10,\,$ and $1000$. Note that
$k>1$ observers fall in from infinity with nonzero initial velocity,
whereas the $k<1$ observers fall from rest at a finite distance.
Similar analyses
have been performed  for several values of $k$ for outgoing geodesic observers
who originate very near the horizon. In all cases, we find that
$|F|(\Delta\tau)^2 \lprox 1$. A corresponding treatment for geodesic observers
moving through the Boulware vacuum yields similar results.

     We now consider four-dimensional black holes. Utilizing numerical
data of Jensen, McLaughlin, and Ottewill \cite{JMO}
for the renormalized quantum stress-tensor of an electromagnetic field in the
Unruh state of a four-dimensional evaporating Schwarzschild black hole,
we have performed a similar analysis for several freely infalling radial
observers in the more limited range from $r=5M$ to $r=2.1M$. The energy
density and flux for the chosen observers are negative throughout this range.
In four-dimensions, the analog of Eq. (~\ref{eq:QI}) is
 \begin{equation}
  |F|A \,(\Delta \tau)^2\lprox 1\,,    \label{eq:4DQI}
\end{equation}
where $A$ is the collecting area for the negative energy.
Here the time interval $\Delta \tau$ will be taken
to be the proper time to fall from $r=5M$ to $r=2.1M$ and $|F|$ will be taken
to be the maximum value of the magnitude of the negative flux, which occurs at
$r=2.1M$. The area $A$ must therefore be less than about $16\pi M^2$.
The following is a representative sample of our numerical results. For $k=10$,
$\Delta\tau=0.29M$ and $(|F|{\Delta\tau}^2)=0.004/{M^2}$;
for $k=1$, $\Delta\tau=3.8M$ and
$(|F|{\Delta\tau}^2)=0.007/{M^2}$; for $k=0.775$, $\Delta\tau=10M$, and
$(|F|{\Delta\tau}^2)=0.03/{M^2}$. The last value of $k$ corresponds to free
fall from rest at $r=5M$. Thus we see that Eq. (~\ref{eq:4DQI}) appears to be
satisfied.

\subsection{Orbiting Observers}
    We now examine geodesic observers in orbit around Schwarzschild and
extreme ($Q=M$) Reissner-Nordstr{\o}m black holes. The line element is
\begin{equation}
ds^2 = -C dt^2 + C^{-1} dr^2 + r^2({d\theta}^2 + {\rm sin}^2{\theta}{d\phi}^2)
\,,\label{eq:4DMetric}
\end{equation}
where $C=(1-2M/r)$ for Schwarzschild and $C=(1-M/r)^2$ for \hbox{$Q=M$}
Reissner-Nordstr{\o}m black holes. The four-velocity for such a
geodesic observer is
\begin{equation}
u^{\mu}=\biggl(1/{\sqrt{(C-rC'/2)}},\, 0,\,
                                   \sqrt{C'/(2rC-r^2C')},\, 0 \biggr)\,.
\label{eq:O4V}
\end{equation}
Since $u^t>0$ and we have chosen $u^{\theta}>0$, we now wish to choose the
observer's normal vector so that $n^{\theta}<0$. The relation
$u^{\mu}n_{\mu}=0$, and the fact that $u^{\mu}$ and $n^{\mu}$ are unit vectors
then imply
\begin{equation}
n^{\mu}=\biggl(-1/{\sqrt{(2C^2/{rC'}-C)}},\, 0,\,
   -\sqrt{2C/[r^2(2C-C'r)]},\, 0 \biggr)\,.
\label{eq:ON}
\end{equation}
For an observer moving
through ``classical dust'', this choice would correspond to a positive
energy flux (i.e., a positive energy flux flowing in the direction opposite
to the direction of the observer's motion). As before, we will regard the
energy flux as negative when the energy density is also negative.
Substitution of Eqs. (~\ref{eq:O4V}) and (~\ref{eq:ON}) into
Eqs. (~\ref{eq:enden}), (~\ref{eq:flux}) yields
\begin{equation}
U={2C\,{T^t}_t-C'\,r\,{T^{\theta}}_{\theta}\over{(C'\,r-2C)}}\,,
\label{eq:4Denden}
\end{equation}
\begin{equation}
F={{\sqrt{2rC\,C'}\, \bigl({T^{\theta}}_{\theta}- {T^t}_t \bigr)}\over
{(2C-rC')}} \,.
\label{eq:4Dflux}
\end{equation}

For a Schwarzschild black hole we have
\begin{equation}
U=(1-3M/r)^{-1}\,\bigl[{T^{\theta}}_{\theta}\,(M/r)-{T^t}_t\,(1-2M/r)
\bigr]\,,
\label{eq:4DendenS}
\end{equation}
\begin{equation}
F={{(M/r)^{1/2}\,(1-2M/r)^{1/2}}\over{(1-3M/r)}}\>
\bigl({T^{\theta}}_{\theta}- {T^t}_t \bigr)\,.
\label{eq:4DfluxS}
\end{equation}
Timelike circular orbits occur for $r>3M$ . In  Ref. \cite {JMO},
the numerical values of
${T^t}_t$ and ${T^{\theta}}_{\theta}$ are plotted for a quantized
electromagnetic field in the Unruh vacuum state in the range from
$r=5M$ to $r=2M$. From Figs. (1) and (3) of Ref. \cite {JMO}, one can see that
${T^{\theta}}_{\theta}<0$ and ${T^t}_t>0$ in the region $3M<r\lprox 5M$. Thus
  from Eqs. (~\ref{eq:4DendenS}) and (~\ref{eq:4DfluxS}), we see
that an orbiting observer who manages to stay in a circular orbit (with small
thrusts to prevent the growth of instabilities) will see $U<0$ and $F<0$.
That is, the observer will see a {\it constant negative} energy flux.
Therefore, it appears that the quantum inequality restrictions on the
flux are violated.

In the case of the evaporating Schwarzschild black hole, it is not clear
whether this violation is due to the existence of the Hawking radiation
or to the motion of the orbiting observer through the static negative energy
background polarization surrounding the hole. It is therefore interesting
to examine a case where the Hawking radiation is absent, such as a $Q=M$
Reissner-Nordstr{\o}m black hole. A quantized field in its vacuum state
on this background possesses more of the properties expected of a ground
state than the same field on the Schwarzschild background.
For this black hole
\begin{equation}
U=-{{\bigl[(r-M)\,{T^t}_t -\,M\,{T^{\theta}}_{\theta}\bigr]}\over{(r-2M)}}\,,
\label{eq:4DendenRN}
\end{equation}
\begin{equation}
F={\sqrt{M(r-M)}\over(r-2M)}\> \bigl({T^{\theta}}_{\theta}- {T^t}_t \bigr)\,.
\label{eq:4DfluxRN}
\end{equation}
In this case, timelike circular orbits occur for $r>2M$.
Anderson, Hiscock, and Samuel \cite{AHS1,AHS2} have recently given a numerical
computation of the stress-energy tensor of quantized scalar fields in
static black hole spacetimes. For $Q=0.99M$, which we take to represent the
extreme $Q=M$ case, they find \cite{AHS2} that
${T^{\theta}}_{\theta}<0$ and ${T^t}_t>0$ in the region $2M<r<4M$.
  From Eqs. (~\ref{eq:4DendenRN}) and (~\ref{eq:4DfluxRN}), we again find
that an orbiting observer in this region will see $U<0$ and $F<0$.
Thus in this case as well, the flux inequality restrictions are violated.

     It is of interest to note that the counterexamples in this subsection
depend upon the ability of the orbiting observer to make many orbits
around the black hole. For example, for the observer orbiting a
Schwarzschild black hole near $r=3M$, one finds that
\begin{equation}
|F|(\Delta \tau)^2 \approx {{0.04 N^2}\over {M^2}},
\end {equation}
where $\Delta \tau$ is the proper time required to complete $N$ orbits.
The collecting area $A$ cannot exceed $M^2$ by a large factor, so for
a single orbit one will have that
 \begin{equation}
  |F|A \,(\Delta \tau)^2 \lprox 1\,.
\end{equation}
A similar result holds in the case of the observers orbiting a
Reissner-Nordstr{\o}m black hole.

\section{Switched Energy Detectors }
\label{sec:detect}

      Particle detector models are a useful probe of the operational
significance of the constructs of quantum field theory. The simplest such
model is the monopole detector \cite{Unruh,DeWitt} in which a quantum system
is coupled linearly to the quantized field. A variation of the monopole
detector was proposed by Grove \cite{Grove} in which the detector is switched
on for only a finite time interval. Switched detectors were further
discussed in Ref. \cite{Svaiter}. Even an inertial detector in the Minkowski
vacuum state will undergo excitations as a result of the switching, and
a flux of negative energy can be interpreted as tending to suppress these
excitations. However, none of these model detectors is really a detector of
the energy density or of other components of the energy momentum tensor.
The response of the detector typically involves a double line integral of
a two-point function over the detector's world line, or a portion of the
world line. In this section, we wish to propose a further modification of
the monopole detector models which will enable one to measure operationally
the expectation value of the stress tensor of a quantized field.

      Let us consider a quantized scalar field $\phi$ which is coupled
to a quantum system (the detector) via the interaction Lagrangian
\begin{equation}
L_I = c \,m(\tau)\, g(\tau) \sum_{j=1}^n \ell_j^\mu \phi_{,\mu}.
\end{equation}
Here $\tau$ is the detector's proper time, $c$ is a coupling constant,
$m(\tau)$ is the monopole moment (an operator with nonvanishing matrix
elements between the different quantum states of the detector), $g(\tau)$
is a switching function which vanishes outside of some finite interval
in $\tau$, and $\bigl\{ \ell_j^\mu \bigr\}$ is a set of vectors. The
analysis of the response of this detector follows the same steps as in
the case of the usual monopole detector \cite{BD2}. The probability for
the detector to be excited from energy eigenstate $|E_0\rangle$ to
energy eigenstate  $|E\rangle$ is found to be
\begin{equation}
P_{E,E_0} = c^2 \Bigl|\langle E| m(0)  |E_0\rangle\Bigr|^2 R(E-E_0),
            \label{eq:prob}
\end{equation}
where the detector's response function, $R$, is given by
\begin{equation}
R(E)= \sum_{i,j=1}^n {\ell_j^\mu} {\ell_i^\nu} \int_{-\infty}^{\infty}
      d\tau d\tau' e^{-iE(\tau-\tau')} g(\tau) g(\tau')
      \langle \phi_{,\mu}(x) \phi_{,\nu}(x') \rangle.
        \label{eq:response}
\end{equation}
Here $x$ is the spacetime location of the detector at time $\tau$, and
$\langle \, \rangle$ denotes the expectation value in $|\psi\rangle$,
the quantum state of the field.

      Let us restrict our attention primarily to flat spacetime models,
in which $|\psi\rangle$ is either a nonvacuum state or a Casimir vacuum state.
We cannot directly measure $R$, but we can measure $R-R_0$, where $R_0$
is the response in the Minkowski vacuum. This
difference is given by
\begin{equation}
R-R_0 = \sum_{i,j=1}^n {\ell_j^\mu} {\ell_i^\nu} \int_{-\infty}^{\infty}
      d\tau d\tau' e^{-iE(\tau-\tau')} g(\tau) g(\tau')
      \langle \phi_{,\mu}(x) \phi_{,\nu}(x')\rangle_R.
        \label{eq:rresponse}
\end{equation}
Here $\langle \, \rangle_R$ is the renormalized expectation value,
the difference
between that in $|\psi\rangle$ and that in the Minkowski vacuum. The
quantity $R-R_0$ is in general a nonlocal quantity which depends upon
field fluctuations over an extended region of spacetime. However, we may
make the switching function $g(\tau)$ sharply peaked in order to probe a
localized  region. This function should satisfy
\begin{equation}
\int_{-\infty}^{\infty} d\tau \, g(\tau) =1.
\end{equation}
Thus in the limit that $g(\tau)$ is sharply peaked at $\tau=\tau_0$,
\begin{equation}
g(\tau) \rightarrow \delta(\tau-\tau_0).
\end{equation}
In this limit, we have
\begin{equation}
R-R_0 = \sum_{i,j=1}^n {\ell_j^\mu} {\ell_i^\nu}
      \langle \phi_{,\mu}(x) \phi_{,\nu}(x) \rangle_R.
        \label{eq:localresponse}
\end{equation}
The quantity $\langle \phi_{,\mu}(x) \phi_{,\nu}(x)\rangle_R$ is finite
and is evaluated at the location of the detector at time $\tau_0$. A
negative value for $\langle \phi_{,\mu}(x) \phi_{,\nu}(x)\rangle_R$
simply means that the detector's response is less than the corresponding
response in the Minkowski vacuum.

      By combining the responses of a set of detectors we may measure
the renormalized expectation value of the stress tensor,
$\langle T_{\mu\nu}(x) \rangle_R$. Consider, for example, the massless
minimally coupled scalar field for which
\begin{equation}
T_{\mu\nu}= \phi_{,\mu} \phi_{,\nu} -
             {1\over 2} g_{\mu\nu}  \phi_{,\rho} \phi^{,\rho}.
\end{equation}
To measure a diagonal component of $\langle T_{\mu\nu}(x) \rangle_R$
in four dimensional spacetime, we need to combine the responses of four
different detectors, each of which measures one diagonal component
of $\langle \phi_{,\mu}(x) \phi_{,\nu}(x) \rangle_R$. That is, to measure
$\langle [\phi_{,t}(x)]^2 \rangle_R$, we take $n=1$ and $\ell^\mu =
t^\mu$, the unit vector in the t-direction. Similarly, we construct
detectors for each of the three spatial directions. To measure an
off-diagonal component, we need three detectors. For example, the
flux in the x-direction is given by the expectation value of
$T_{tx} = {1\over 2} (\phi_{,t}\, \phi_{,x} +\phi_{,x}\, \phi_{,t})$.
We take the first detector to be
specified by $n=2$ and ${\ell_1^\mu}= t^\mu$ and ${\ell_2^\mu}= x^\mu$,
where $ x^\mu$ is the unit vector in the x-direction. Equation
(~\ref{eq:localresponse}) for this detector becomes
\begin{equation}
R-R_0 = \langle [\phi_{,t}(x)]^2 + \phi_{,t}(x) \phi_{,x}(x)
      + \phi_{,x}(x) \phi_{,t}(x)  +[\phi_{,x}(x)]^2 \rangle_R.
\end{equation}
By combining this with the response of detectors which measure
$\langle [\phi_{,t}]^2 \rangle_R$ and $\langle [\phi_{,x}]^2 \rangle_R$,
we obtain $\langle T_{tx} \rangle_R$.

     The various detectors which are required to measure a particular
component of $\langle T_{\mu\nu}(x)\rangle_R$ must all make measurements
in a spacetime region which is small compared to the scales over which
$\langle T_{\mu\nu}(x) \rangle_R$ varies. One might be concerned that the
different detectors could disturb one another. However, this effect may be
made small by choosing the coupling constant $c$ to be sufficiently small.
  From Eq. (~\ref{eq:prob}), we see that the effect of the quantized field
upon any individual detector is of order $c^2$. The different detectors
are assumed to interact with one another only through their coupling to the
field. Thus the disturbance in one detector due to another detector is
necessarily of higher order in $c$ and can be minimized by making $c$
small.

     The main result of this section is that we have given a prescription
by which the stress tensor of a quantized scalar field may be
measured by non-gravitational means. Note that this prescription answers
a challenge proposed by Padmanaban and Singh \cite{PS}, although the
method we have used is rather different from that envisioned by these authors.
We have restricted our attention to flat spacetimes, but this procedure
could be generalized to curved spacetime. In order to do this, one would
have to modify the definition of
$\langle \phi_{,\mu}(x) \phi_{,\nu}(x) \rangle_R$ to include subtraction
of the curvature-dependent divergences which arise in a nonflat spacetime.

     Although the above model illustrates the possibility of constructing
a non-gravitational stress-energy detector, we saw that measurement of
$\langle T_{\mu\nu}(x)\rangle$ at a spacetime point necessarily entailed
rapid switching of the detector. As a result, such a detector may tell
us little about the (possibly large) cumulative effects of negative energy.
An example of dramatic cumulative effects of negative energy is black hole
evaporation. Conversely, a detector which is not rapidly switched involves
an integral over the detector's world line and hence measures a nonlocal
quantity, rather than $\langle T_{\mu\nu}(x)\rangle$.

\section{Summary and Discussion}
\label{sec:summary}

      In this paper, we have been concerned with effects due to
the motion of an observer through a stationary distribution of negative
energy, in contrast to dynamically generated negative energy fluxes
of the sort which are created by quantum coherence effects in Minkowski
spacetime. One of our goals has been to test the extent to which quantum
inequalities like Eq. (~\ref{eq:QI}) are satisfied by the fluxes due to
such motion. We have found that these inequalities appear to be satisfied
for geodesic observers moving radially in two and four dimensional black
hole spacetimes. However, they are not satisfied in all cases. We found
that observers who move sufficiently slowly through the Casimir vacuum
may violate the inequality arbitrarily. This is likely to be a general
feature of slow motion through static negative energy regions.
Similarly, observers in
circular orbits around black holes may see a constant negative energy
flux in their frame of reference. (Note that in the latter case, the
observer's velocity does not play a crucial role, as it is determined
by the orbit and is not a free parameter.) In both cases, we have
examples of inertial observers who see negative energy fluxes which are
not constrained by a quantum inequality such as Eq. (~\ref{eq:QI}),
which limits the magnitude and duration of the flux.

     The physical implications of these counterexamples are still
somewhat obscure.
It is not clear whether the effective negative energy flux produced
by motion relative to a stationary negative energy distribution can
be absorbed to create macroscopic effects. There may be fundamental
differences between these quantum states and those involving negative
energy fluxes produced by quantum coherence effects. In order to address this
issue, one would want to construct model energy detectors. Most
quantum detectors do not couple to the local energy density \cite{PS}.
We have seen that non-gravitational stress-energy detectors can be devised,
but they appear to require rapid switching. This prevents them from
registering cumulative effects of negative energy. Thus it remains an
open question as to whether negative energy fluxes can be manipulated
to produce gross macroscopic effects.

\vskip 0.2 in
\centerline{\bf Acknowledgements}
We would like to thank Bruce Jensen, Bernard Kay, Nami Svaiter
and Kip Thorne for
helpful discussions and Paul Anderson for providing unpublished results
of numerical calculations of the stress tensor in the Reissner-Nordstr{\o}m
spacetime.  This research was supported in part by NSF Grant
No. PHY-9208805 and by an AAUP/CCSU Faculty Research Grant.

\newpage

\section*{Figure Captions}
\begin{itemize}

\item{[1]}  Graph of the magnitude, $|F|$, of the negative energy flux
in units of $M^{-2}$ as a function of proper time $\tau$ in units of $M$.
Here the observer falls inward from rest at infinity, i.e. $k=1$. The
flux diverges as the observer approaches the singularity at $r=0$.
In this region, the
magnitude of the flux doubles on a timescale of order $\Delta\tau
\lprox 0.01 M$.

\end{itemize}

\end{document}